\documentclass[amsmath, amssymb,10pt,aps,prl,reprint,notitlepage,showpacs,superscriptaddress,longbibliography]{revtex4-1}
\usepackage{graphicx}
\usepackage{calc}
\usepackage{braket}
\usepackage{ulem}
\usepackage{slashed}
\usepackage{wasysym}
\usepackage{siunitx}
\usepackage{dsfont}
\usepackage{amsthm,amsmath,amsfonts,amssymb,verbatim,color}
\usepackage{graphicx}
\usepackage{subfigure}
\usepackage{bm}
\usepackage{epsfig,slashed}
\usepackage[T1]{fontenc}
\usepackage{ulem}
\usepackage[colorlinks=true,citecolor=blue,linkcolor=blue,urlcolor=blue]{hyperref}
\begin{document}
\title{Direct imaging of antiferromagnetic domains and anomalous layer-dependent mirror symmetry breaking in atomically thin  MnPS$_3$}

\author{Zhuoliang Ni}
\affiliation{Department of Physics and Astronomy, University of Pennsylvania, Philadelphia, Pennsylvania 19104, U.S.A}
\author{Huiqin Zhang}
\affiliation{Department of Electrical and System Engineering, University of Pennsylvania, Philadelphia, Pennsylvania 19104, U.S.A}
\author{David Hopper}
\affiliation{Department of Physics and Astronomy, University of Pennsylvania, Philadelphia, Pennsylvania 19104, U.S.A}
\affiliation{Department of Electrical and System Engineering, University of Pennsylvania, Philadelphia, Pennsylvania 19104, U.S.A}
\author{Amanda V. Haglund}
\affiliation{Department of Materials Science and Engineering, University of Tennessee, Knoxville, TN 37996, U.S.A.}
\author{Nan Huang}
\affiliation{Department of Materials Science and Engineering, University of Tennessee, Knoxville, TN 37996, U.S.A.}
\author{Deep Jariwala}
\affiliation{Department of Electrical and System Engineering, University of Pennsylvania, Philadelphia, Pennsylvania 19104, U.S.A}
\author{Lee Bassett}
\affiliation{Department of Electrical and System Engineering, University of Pennsylvania, Philadelphia, Pennsylvania 19104, U.S.A}
\author{David G. Mandrus}
\affiliation{Department of Materials Science and Engineering, University of Tennessee, Knoxville, TN 37996, U.S.A.}
\affiliation{Materials Science and Technology Division, Oak Ridge National Laboratory, Oak Ridge, TN, 37831, U.S.A.}
\author{Eugene J. Mele}
\affiliation{Department of Physics and Astronomy, University of Pennsylvania, Philadelphia, Pennsylvania 19104, U.S.A}
\author{Charles L. Kane}
\affiliation{Department of Physics and Astronomy, University of Pennsylvania, Philadelphia, Pennsylvania 19104, U.S.A}
\author{Liang Wu}
\email{liangwu@sas.upenn.edu}
\affiliation{Department of Physics and Astronomy, University of Pennsylvania, Philadelphia, Pennsylvania 19104, U.S.A}

\date{\today}

\begin{abstract}

We have developed a sensitive cryogenic second-harmonic generation microscopy to study a van der Waals antiferromagnet MnPS$_3$. We find that long-range N\'eel antiferromagnetic order develops from the bulk crystal down to the bilayer, while it is absent in the monolayer.   Before entering the long-range antiferromagnetic ordered phase in all samples, an upturn of the second harmonic generation below 200 K indicates the formation of the short-range order and magneto-elastic coupling.  We also directly image the two antiphase (180$^{\circ}$) antiferromagnetic domains and thermally-induced domain switching down to bilayer.   An anomalous mirror symmetry breaking shows up in samples thinner than ten layers for the temperature both above and below the N\'eel temperature, which indicates a structural change in few-layer samples. Minimal change of the second harmonic generation polar patterns in strain tuning experiments indicate that the symmetry crossover at ten layers is most likely an intrinsic property of MnPS$_3$ instead of an extrinsic origin of substrate-induced strain.  Our results show that second harmonic generation microscopy is a direct tool for studying antiferromagnetic domains in atomically thin materials, and opens a new way to study two-dimensional antiferromagnets.
\end{abstract}

\maketitle
Compared to ferromagnetic materials, antiferromagnetic (AFM) materials usually have a much higher magnon frequency in the terahertz regime and a more stable ground state to perturbations of external magnetic fields, making them  promising platforms for  spintronic devices operating in the terahertz frequency range. Two-dimensional (2D) AFM materials are of particular interest recently \cite{leenanoletter16,wang2dmat16,kimprl18, sunjpcl19,Kim_2019,kimnatcomm19, XingPRX2019,longnanoletter20,kangnature20,vsivskinsnatcomm20,vaclavkova2dmater20,wangnatmater21,ninatnano21,hwangbonatnano21,Afanasievsciadv21}, with the prospect of  devices down to the atomically-thin limit. Contrary to the rapid development in 2D  ferromagnetic  materials \cite{huangnat17,gongnat17,Chensci19,thielsci19,kleinsci18,songsci18,burchnat18,huangnatnano18,dengnat18,maknatrevphys19,gongsci19,gibertininatnano19,sunnat19},  2D AFM  materials have been less explored due to the lack of direct probes on the AFM orders until very recently \cite{chuprl20,kangnature20,hwangbonatnano21,ninatnano21,wangnatmater21}. Raman scattering \cite{leenanoletter16,wang2dmat16, sunjpcl19,Kim_2019,mccrearyprb2020,mai2020magnon}, magnetic tunneling \cite{longnanoletter20} and magnon transport measurement reveal photons or the bandgap coupled to the AFM orders \cite{XingPRX2019}, but they do not directly probe the AFM order and the domains \cite{maknatrevphys19}.

Commonly known as a probe of structural and surface inversion symmetry ($\mathcal{P}$) breaking, second harmonic generation (SHG) has also been demonstrated to be a direct probe to the long-range AFM orders and domains \cite{fiebigprl94,Fiebigjosab05,sunnat19,chuprl20,ninatnano21}. SHG can be classified into two kinds: i-type and c-type\cite{fiebigprl94,Fiebigjosab05}. The i-type SHG is invariant under the time-reversal symmetry ($\mathcal{T}$), and includes the electric-dipole (ED) SHG in noncentrosymmetric crystals and electric-quadruple (EQ) SHG. The EQ contribution does not require the condition of broken inversion symmetry in the lattice. The c-type SHG changes sign under the time-reversal transformation, and is only present if the spin structure breaks the inversion symmetry in centrosymmetric lattice. The c-type SHG has been very rare, and was observed in parity-time-reversal  ($\mathcal{PT}$) symmetric antiferromagnets such as Cr$_2$O$_{3}$ \cite{fiebigprl94,Fiebigjosab05}, bilayer CrI$_3$ \cite{sunnat19} and monolayer MnPSe$_3$ \cite{ninatnano21}. Below the transition temperature, the c-type SHG susceptibility is proportional to the order parameter \cite{muthukumarprl95}. The interference between the c-type and i-type SHG has been used to directly image AFM domains \cite{fiebigprl94,Fiebigjosab05, ninatnano21}. Nevertheless, studies on imaging of AFM domains in atomically thin  crystals have been very limited \cite{sunnat19,ninatnano21}.

\begin{figure}
\centering
\includegraphics[width=0.5\textwidth]{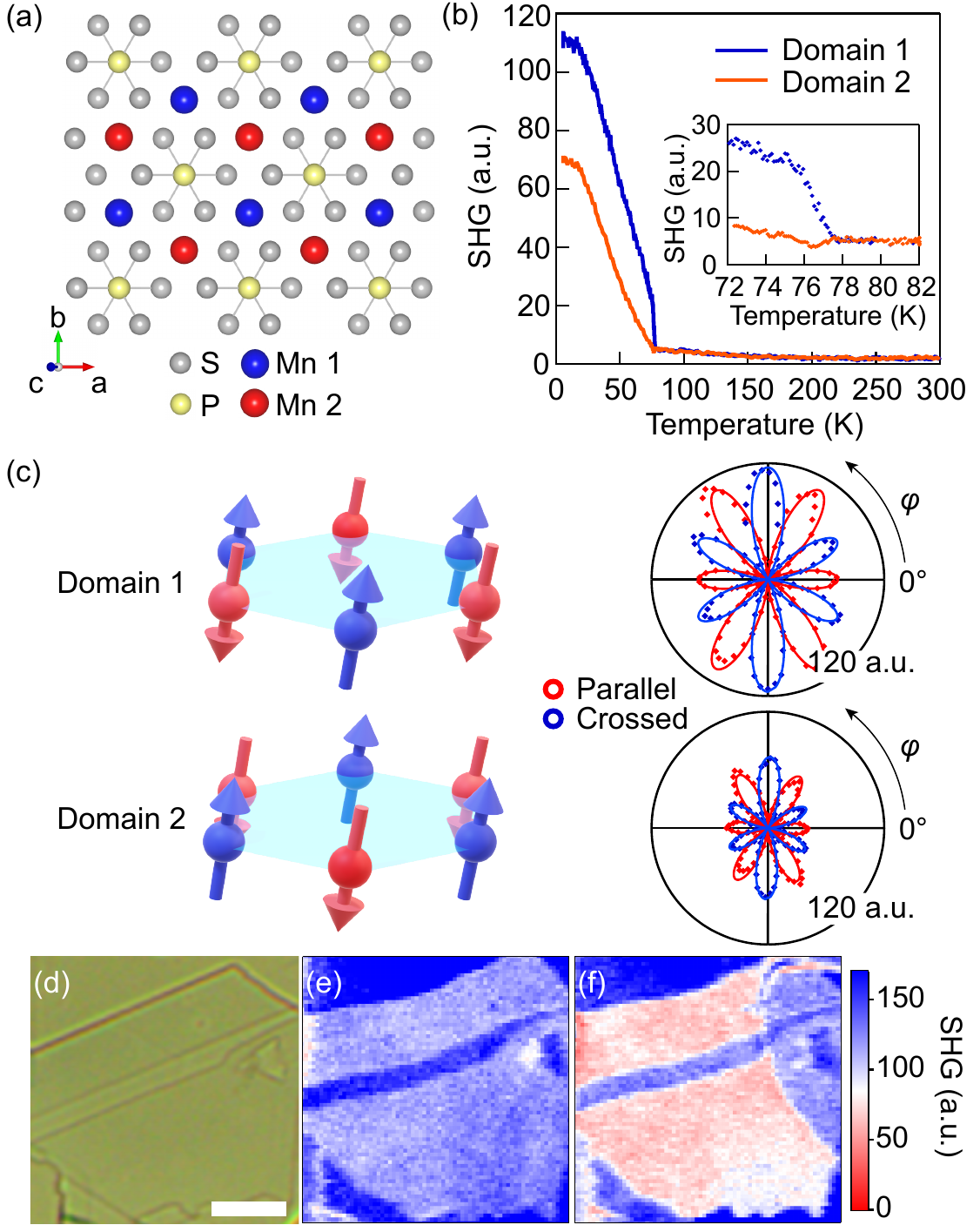}
\caption{ (a) Top view of the monolayer crystal structure of MnPS$_3$. (b) Temperature dependence of SHG intensity of two domains in the same MnPS$_3$ bulk crystal. The inset shows SHG intensity near the transition temperature. (c) Two possible spin configurations of Mn atoms in a single-layer. Red arrows indicate spin directions of Mn atoms.  The SHG polar patterns corresponding to two domains are shown on the right. The measured data are shown in dots and best fits are shown in solid lines. (d)-(f) Optical image and SHG intensity images at 5 K of a bulk MnPS$_3$ sample.  Scale bar: 50 $\mu$m. }
\label{fig1}
\end{figure}

The AFM transition metal thiophosphates MPS$_3$ (M=Mn, Fe, Ni) have attracted lots of interests recently  including the many-body exciton in NiPS$_3$ \cite{kangnature20,hwangbonatnano21,wangnatmater21}, Ising antiferromagntism in FePS$_3$ \cite{scagliottiprb1987,leenanoletter16,wang2dmat16,mccrearyprb2020}, and magneto-electric effect in MnPS$_3$ \cite{ressoucheprb10}. Also, they order in different spin structures despite the same monoclinic structure in space group No. 12. NiPS$_3$ and FePS$_3$ have zigzag orders, but  MnPS$_3$ has the N\'eel order. Among them, MnPS$_3$ breaks $\mathcal{P}$ and $\mathcal{T}$ below the N\'eel temperature ($T_N$), but respects $\mathcal{PT}$ symmetry, and, therefore, it allows the c-type ED  SHG in the AFM phase. A previous SHG study on MnPS$_3$  demonstrated a phase-transition-like feature around $T_N$ down to seven-layer (7L), but did not observe short-range order above $T_N$, N\'eel-type AFM domains, and did not provide a clear evidence for the c-type SHG that changes sign under time-reversal transformation \cite{chuprl20}. Very high laser power was used in the previous study \cite{chuprl20}, because of the material's weak second-order response. Note that the SHG intensity from MnPS$_3$ is at least two orders of magnitude lower than that in MnPSe$_3$ due to its much weaker spin-orbital coupling \cite{ninatnano21}.  Fast sample degradation due to high laser power happened within hours, which prevent a systematic study of the layer-dependent antiferromagnetism, domains, short-range order and time-reversal-odd c-type SHG in MnPS$_3$ \cite{chuprl20}.

In this work, we develop a  sensitive cryogenic scanning SHG microscopy with the photon detection sensitivity of 0.1 counts per second (c.p.s.) by using a photon counter to systematically study the layer-dependent antiferromagnetism in MnPS$_3$ down to the monolayer. We observe the sign change  of the c-type SHG coefficient below $T_N$, and two antiphase (180 $^{\circ}$) AFM domains with spins reversed by time-reversal transformation. The long-range AFM order, evident from the c-type SHG, is present from the bulk to bilayer, but is absent in the monolayer.  Between $T_N$  and 200 K, we observe an increase of the SHG as a function of decreasing temperature, which indicates correlation-induced short-range order and the magneto-elastic coupling. Polarized SHG measurement shows the mirror-symmetry breaking below 10L both below and above $T_N$, indicating a symmetry crossover in the structure in thin flakes.

The bulk crystal of MnPS$_3$ belongs to the point group of 2/m above $T_N$ $\approx$ 78 K and magnetic point group 2'/m below $T_N$. Neutron scattering measurements indicate its magnetic system to be Heisenberg type with tiny dipolar anisotropy and single-ion anisotropy  \cite{ronnowPhysicaB2000, wildesprb2006,ressoucheprb10,okudajpsj86,pichJMM1995}. The lattice structure is shown in Fig. \ref{fig1}(a). In the single layer, Mn atoms form a honeycomb lattice in the $ab$ plane, and the two Mn atoms with opposite spin directions are marked by different colors, showing its  AFM N\'eel order in a single layer. The spins are mainly out of plane, but 8 degree from the surface normal direction. The honeycomb layers stack along the $c$ axis and with a shift of 1/3 of the lattice constant along the $a$ axis, which gives rise to a centrosymmetric lattice. There is a mirror symmetry perpendicular to the $b$ axis.   Interestingly, the formation of N\'eel AFM order breaks the inversion symmetry, and therefore it allows c-type ED SHG.  

We first explore the SHG signal in a bulk crystal with thickness around 10 $\mu$m as shown in Fig. \ref{fig1}(d). Under normal incidence, we send in laser pulses centered at 800 nm (with 80 MHz repetition rate and around 50 fs duration) and measure SHG at 400 nm \cite{wunatphys17,ninatnano21}. We focus 5 mW on a spot with a diameter of 10 $\mu$m. By consecutively doing thermal cycles at one spot, Fig. \ref{fig1}(b) shows that there are only two  SHG traces when crystal is cooled from the paramagnetic to the AFM states\cite{sm}. At high temperatures from 200 K to 300 K, the temperature-independent SHG signal contributes from the electric-quadruple (EQ) term of the lattice. From 200 K to 78 K ($T_N$), the two curves still overlap and rise gradually. The slow rising of EQ term in SHG is believed to originate from short-range orders of spins  due to  correlation-induced lattice distortions \cite{wildesprb2006}, which is also known as magneto-elastic coupling \cite{vaclavkova2dmater20}.  A similar phenomenon was observed previously in a van-der Waals ferromagnet \cite{ronnatcomm19}. The two curves split at 77.7($\pm$0.3) K, which we define as the onset of c-type ED SHG and therefore the $T_N$.   To explain the transition behavior, we write the interfering SHG intensity from ED and EQ terms as
\begin{equation}
    I_{i}^{SHG}=\left(\pm  \chi_{ijk}^{ED} (T \leq T_N)  E_{j}  E_{k} +\chi_{ijkl}^{EQ}(T)  E_{j} \nabla_{k} E_{l}\right)^2,
\end{equation}
 where $\chi^{ED}_{ijk}(T) \propto$ $M$ $\propto(T_N-T)^{\beta}$ is the ED SHG coefficient \cite{muthukumarprl95}. ($M$ is the staggered magnetization of the two Mn sites and is the order parameter.)  The $\pm$ sign corresponding to two  $\mathcal{PT}$-related domains as shown in \ref{fig1}(c). $\chi^{EQ}(T)$ is present at all temperature and is identical in both domains, since magneto-elastic distortions caused by the two domains are the same.  The splitting of SHG intensity of the two domains at $T_N$ is the direct evidence of the time-reversal-odd c-type ED SHG and therefore the long-range AFM order in MnPS$_3$. As observed in a centrosymmetric 2D ferromagnet \cite{ronnatcomm19} and the 2D zigzag antiferromagnet NiPS$_3$\cite{sm}, just observing the rising of SHG across $T_N$ without observing a sign change could contribute from i-type SHG due to  magneto-elastic coupling and might not be proportional to the order parameter. Note that the inset figure in Fig. \ref{fig1}(b) shows a change of slope in the curve of domain 1 around 76 K, which might be related with the crossover of 3D to 2D critical behaviour \cite{wildesprb2006, sm}.

In Fig. \ref{fig1}(e,f), we show SHG intensity mapping at 5 K after two consecutive thermal cycles in the area marked by the optical image in Fig. \ref{fig1}(d). Fig. \ref{fig1}(e) shows a single domain, while Fig. \ref{fig1}(f) shows two domains. To the best of our knowledge, this is the first report of direct observation of AFM domains in MnPS$_3$. We further measure the SHG polar patterns in these two domains as shown on the right of Fig. \ref{fig1}(b). Here ``parallel'' and ``crossed'' refer to configurations of the polarization of the incident and SHG laser pulses\cite{wunatphys17, ninatnano21}, and they have different angle dependence on $\chi^{ED}_{ijk}$, which reveal the symmetry of the material (see supplementary note S1). The $\phi=0$ direction in parallel configuration corresponds to the $a$-axis of the MnPS$_3$ crystal. These patterns can be well fitted (solid lines) by the symmetry analysis based on the material's space group . The SHG maps in Fig. \ref{fig1}(e,f) and traces in Fig. \ref{fig1}(b) are taken when $\phi=0 ^\circ$ in the parallel configuration. We fix the laser position at one spot, and confirm that there are only two kinds of polar patterns after ten thermal cycles.  The two domains switch to each other by the time-reversal transformation. As a result, the c-type ED SHG is also called non-reciprocal SHG. The polar patterns also confirm that the mirror plane perpendicular to the $b$ axis and the tilting angle of spins are in the $ac$ plane.

 \begin{figure}
\centering
\includegraphics[width=0.5\textwidth]{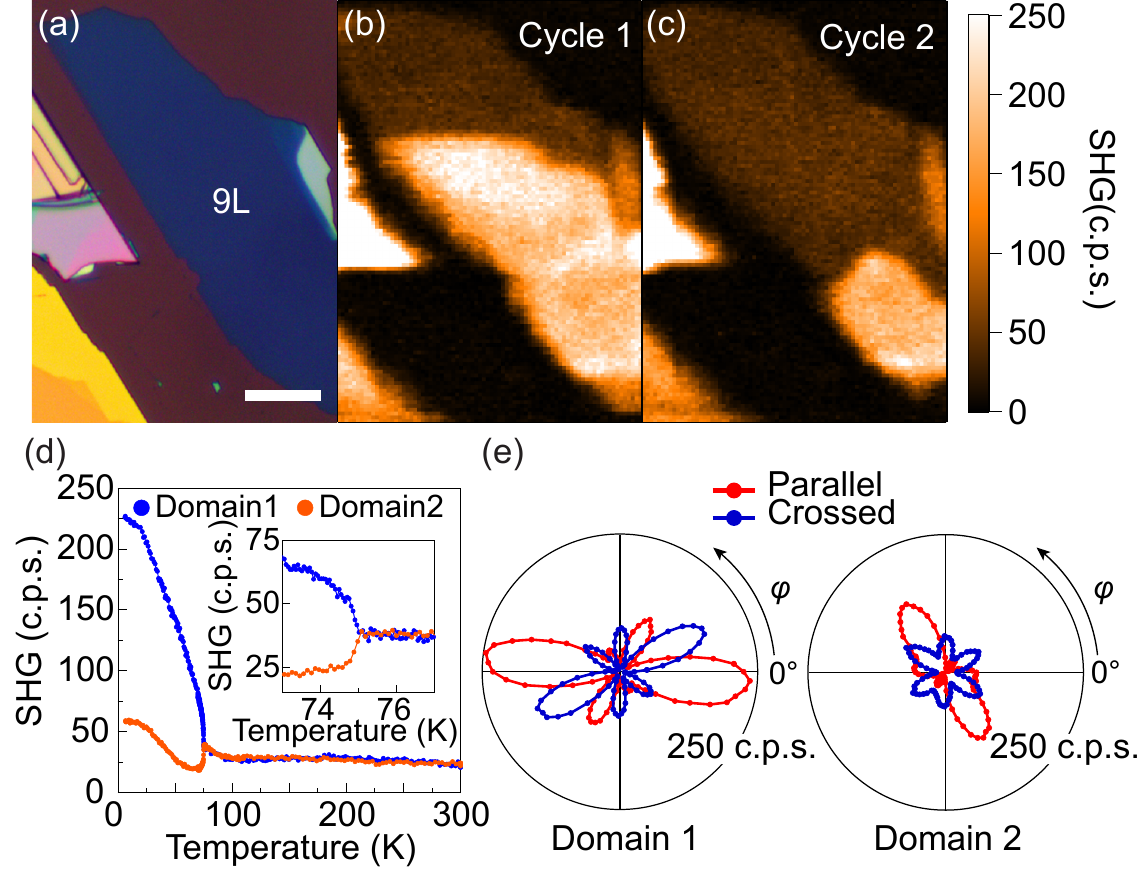}
\caption{ (a) Optical image of the 9L MnPS$_3$ sample. Scale bar: 20 $\mu$m. (b-c) SHG intensity mapping measured at 5 K.  (d) Temperature dependence of SHG intensity of two domains  Inset shows a closer look near transition temperature. (e) Polarization dependence of SHG intensity of two domains measured at 5 K.}
\label{fig2}
\end{figure}

Next, we study whether the long-range order and domains sustain in the atomically thin limit. We use a relatively low power of 2 mW over the beam diameter of 10 $\mu$m to avoid damage to atomically-thin samples. Note that our incident power per area is around 50 times lower than that of the previous study \cite{chuprl20}.  The optical image of a large nine-layer (9L) sample on SiO$_2$(90 nm)/Si(0.5 mm) is shown Fig. \ref{fig2}(a), and the thickness is determined by a combination of atomic force microscopy measurement and color contrast measurement\cite{sm}. We perform a SHG mapping at 5 K (Fig. \ref{fig2}(b)) with the polarization at 172$^\circ$ peak position in the parallel configuration of domain 1 in Fig. \ref{fig2}(e). Two AFM domains are observed represented by the darker (domain 2) and the brighter (domain 1) areas. Next, we warm up the sample above $T_N$ and cool it back to 5 K, and the new SHG mapping is shown as Fig. \ref{fig2}(c). The different distribution of two domains indicates that N\'eel vectors switch freely between two directions across $T_N$, in consistency with the bulk. Fig. \ref{fig2}(d) shows the temperature-dependent intensity of two domains with $\phi$=172$^\circ$ in the parallel configuration. We also perform over nine thermal cycles and confirm that there are only two domains\cite{sm}.  Surprisingly, both polar patterns in  Fig. \ref{fig2}(e) show the breaking of mirror symmetry in the spin structure of this 9L sample.

\begin{figure*}
\centering
\includegraphics[width=0.9\textwidth]{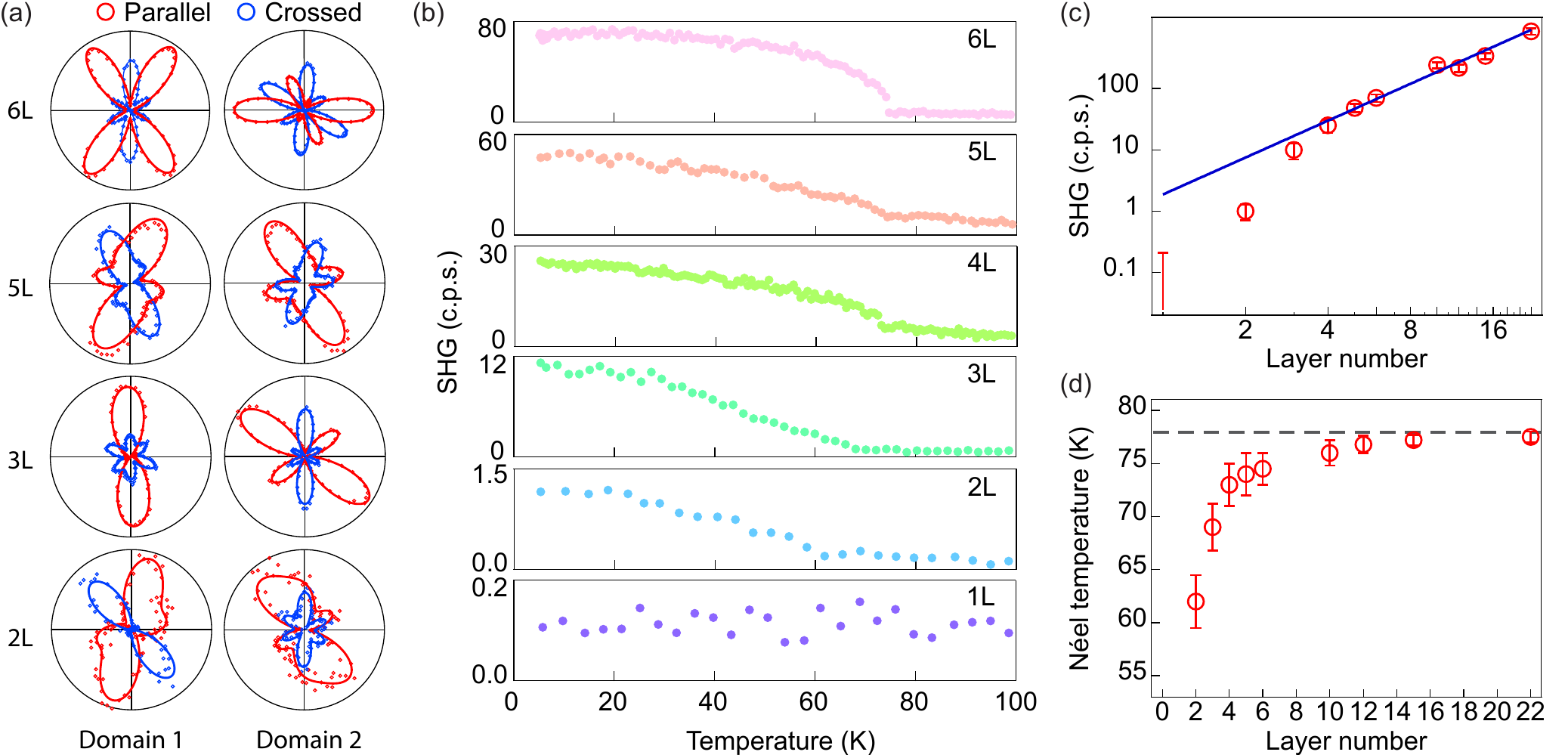}
\caption{ (a) SHG polar patterns of two AF domains in different layers measured at 5 K. (b) Temperature dependence of SHG intensity at the crossed peak of domain 1 from different layers. (c) Layer number dependence of SHG intensity.  (d) Layer number dependence of N\'eel temperature.}
\label{fig3}
\end{figure*}

To approach the 2D limit, we exfoliate MnPS$_3$ samples from six-layer to monolayer. The isolated samples with diameters at least three times larger than laser spot are chosen. The experimental results are summarized in Fig. \ref{fig3}. Polar patterns of the two domains at 5 K in 2,3,5,6-layer samples are shown in Fig. \ref{fig3}(a).  Similar to the 9L sample, the mirror symmetry is broken in these samples. The temperature-dependent SHG intensity of these samples is shown in Fig. \ref{fig3}(b). The layer-dependent intensity at 5 K (chosen at the maximum in the parallel pattern) is shown in Fig. \ref{fig3}(c). As each MnPS$_3$ layer produces in-phase SHG response, the intensity of samples from 3L to 22L nearly follows the dependence of the square of the number of layers, as observed in other inversion-breaking 2D materials \cite{ninatnano21,zhaolight16}. There is a strong suppression of SHG intensity in the bilayer samples. We confirm that the suppression is not due to degration in air as the SHG intensity of a 2L sample exfoliated in a glove box is the same\cite{sm}.  We measure three monolayer samples exfoliated in air and one monolayer sample exfoliated in the glove box, but we are unable to detect a temperature-dependent SHG signal within our setup sensitivity of 0.1 c.p.s.\cite{sm}. Therefore, we think the absence of long-range order in the monolayer is not due to air degration.  The N\'eel temperature, which we extract at the splitting of SHG of two domains, is plotted in  Fig. \ref{fig3}(d).  It is also suppressed with decreasing of the layer number(below 10L).  

\begin{figure}
\centering
\includegraphics[width=0.5\textwidth]{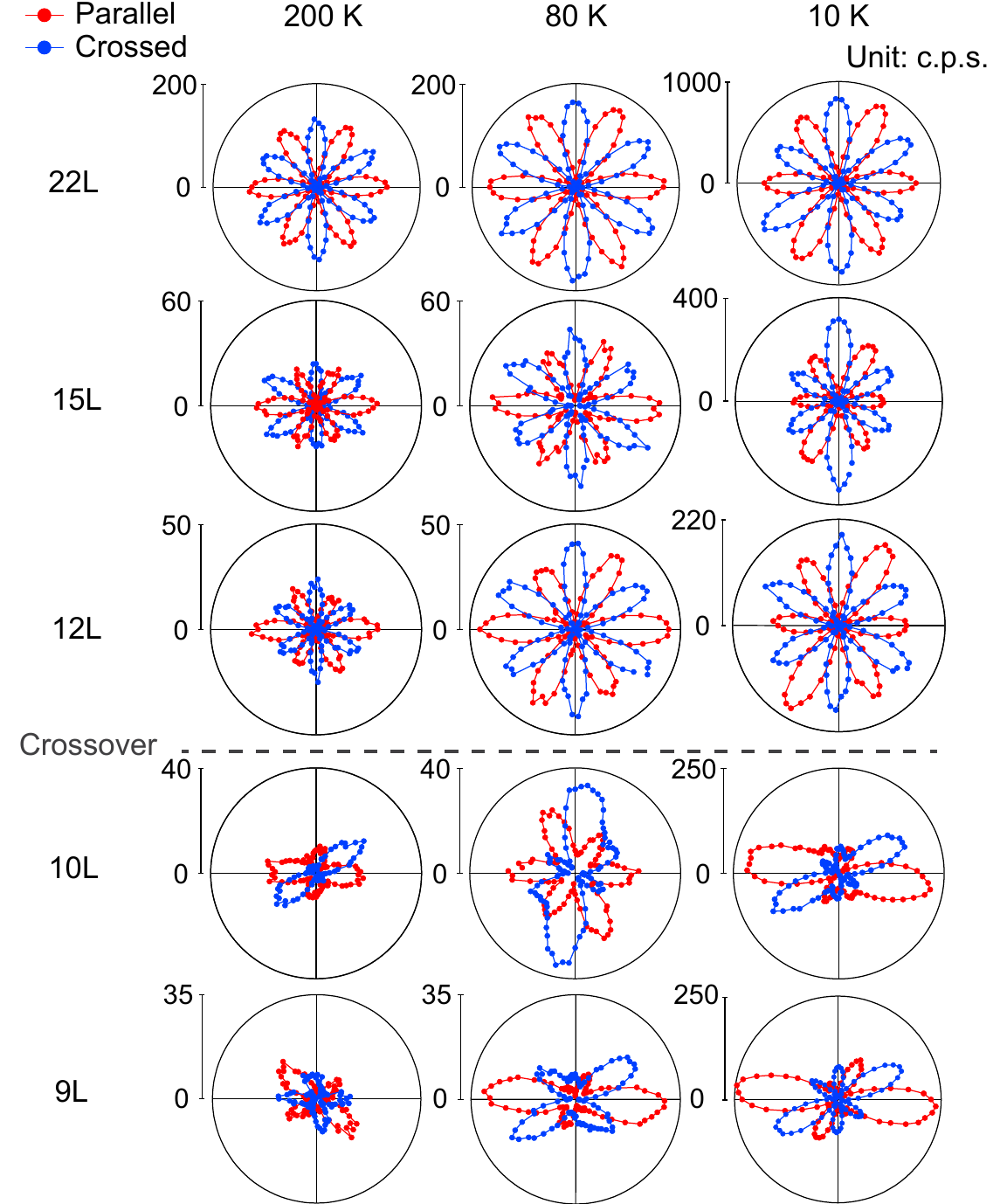}
\caption{ Parallel (blue) and crossed (red) polar patterns of different layers.  The dots with lines are the data.}
\label{fig4}
\end{figure}

In order to study the origin of the mirror symmetry breaking, we systematically investigate the layer dependence. As shown in Fig. \ref{fig4}, there is a clear difference between the 12L and the 10L samples. For samples thicker than 12L, the SHG patterns show a mirror symmetry, and the six lobes are of comparable magnitude, which is similar to the bulk sample in Fig. \ref{fig1}. However, for those thinner than 10L, the mirror symmetry is absent in 10 K, similar to the data shown in Fig. \ref{fig2} and Fig. \ref{fig3}. Note that at 200 K, where the EQ SHG signal is mostly from the lattice and is decoupled to the short-range spin correlation, the mirror symmetry is still absent. Therefore,  the absence of the mirror symmetry below 10L has a lattice origin. A recently near-field infrared spectroscopy study on the phonon spectrum and density functional theory calculation on the same material also shows a crossover of the lattice symmetry around 10L, and they attribute it to the symmetry crossover from 2/m to a higher symmetry P3$_1$m with three mirrors \cite{nealprb2019,nealprb2020}. In our experiment, though observing the similar thickness for the crossover, we find the symmetry of  MnPS$_3$ samples thinner than 10L in the paramagnetic state is actually lower than thicker samples.  We tend to believe that the mirror symmetry breaking is an intrinsic property of very thin MnPS$_3$ crystal as the change of polar patterns between 12L and 10L is dramatic. We tend to think that mirror symmetry breaking is not due to stacking faults as it has a systematic thickness dependence. Also, to exclude the strain effect caused by the substrate, we perform the strain-tuning experiments on a 9L and a 16L samples. No strain-induced mirror symmetry breaking is observed in the 16L sample. Also, both samples do not show the change in the SHG polar pattern by the 5 $\%$ strain applied on the substrate\cite{sm}.  Further detailed structure measurement and density functional theory calculation are called to reveal the origin of mirror symmetry breaking below 10L. 

The long-range order in bulk MnPS$_3$ is believed to result from  the anisotropy and the weak interplanar coupling \cite{ronnowPhysicaB2000,wildesprb2006}. The suppression of the N\'eel temperature possibly results from reduced interplanar coupling in atomically thin samples \cite{kimnatcomm19,gongnat17}. Nevertheless, consider the symmetry of the spin structure is different in samples less than 10L, it might not be valid to use the same bulk Hamiltonian to explain atomically thin samples. We also can not rule out the possibility of disorder effect on the absence of the ordering in the monolayer. More discussions about the suppression of long-range order in the monolayer can be found in Ref.\cite{sm}. We hope more future works will investigate the monolayer MnPS$_3$.

To summarize, by using a  sensitive SHG microscopy, we systematically study the layer dependence of AFM states in MnPS$_3$, and observe the switching of N\'eel-type AFM domains.   Looking forward, We believe that this work is critical to understanding the 2D antiferromagnetism in this compound, and application for 2D terahertz AFM spintronic devices.   \\

\begin{acknowledgments}
We thank S.W. Cheong for helpful discussions. The project design, data collection and analysis are supported by L.W.'s startup package at the University of Pennsylvania. The development of the SHG photon counter is supported by the ARO YIP award under the Grant W911NF1910342. The measurement by the atomic force microscopy is supported by the ARO MURI  under the Grant W911NF2020166. Z.N., H.Z., D.J. and L.W.  acknowledge partial support from National Science Foundation supported University of Pennsylvania
Materials Research Science and Engineering Center (MRSEC)(DMR-1720530). E.J.M. acknowledges support from NSF EAGER 1838456. C.L.K is supported by a Simons Investigator grant from the Simons Foundation. D.G.M acknowledges support from the Gordon and Betty Moore Foundation’s EPiQS Initiative, Grant GBMF9069.  Z.N. and H.Z. also acknowledge support from Vagelos Institute of Energy  Science  and  Technology  graduate  fellowship  at  the  University  of  Pennsylvania.
\end{acknowledgments}

\end{document}

\title{Supplementary Material for\\
\normalsize{Direct imaging of antiferromagnetic domains and anomalous layer-dependent mirror symmetry breaking in atomically thin  MnPS$_3$}}

\author{Zhuoliang Ni}
\affiliation{Department of Physics and Astronomy, University of Pennsylvania, Philadelphia, Pennsylvania 19104, U.S.A}
\author{Huiqin Zhang}
\affiliation{Department of Electrical and System Engineering, University of Pennsylvania, Philadelphia, Pennsylvania 19104, U.S.A}
\author{David Hopper}
\affiliation{Department of Electrical and System Engineering, University of Pennsylvania, Philadelphia, Pennsylvania 19104, U.S.A}
\affiliation{Department of Physics and Astronomy, University of Pennsylvania, Philadelphia, Pennsylvania 19104, U.S.A}
\author{Amanda V. Haglund}
\affiliation{Department of Materials Science and Engineering, University of Tennessee, Knoxville, TN 37996, U.S.A.}
\author{Nan Huang}
\affiliation{Department of Materials Science and Engineering, University of Tennessee, Knoxville, TN 37996, U.S.A.}
\author{Deep Jariwala}
\affiliation{Department of Electrical and System Engineering, University of Pennsylvania, Philadelphia, Pennsylvania 19104, U.S.A}
\author{Lee Bassett}
\affiliation{Department of Electrical and System Engineering, University of Pennsylvania, Philadelphia, Pennsylvania 19104, U.S.A}
\author{David G. Mandrus}
\affiliation{Department of Materials Science and Engineering, University of Tennessee, Knoxville, TN 37996, U.S.A.}
\affiliation{Materials Science and Technology Division, Oak Ridge National Laboratory, Oak Ridge, TN, 37831, U.S.A.}
\author{Eugene J. Mele}
\affiliation{Department of Physics and Astronomy, University of Pennsylvania, Philadelphia, Pennsylvania 19104, U.S.A}
\author{Charles L. Kane}
\affiliation{Department of Physics and Astronomy, University of Pennsylvania, Philadelphia, Pennsylvania 19104, U.S.A}
\author{Liang Wu}
\email{liangwu@sas.upenn.edu}
\affiliation{Department of Physics and Astronomy, University of Pennsylvania, Philadelphia, Pennsylvania 19104, U.S.A}

\maketitle

\section{Symmetry information of the polarization-resolved SHG intensity}
In this study, we mainly focus on two second-harmonic contributions in MnPS$_3$: the spin-induced electric dipole (ED) term ($\chi^{ED}_{ijk}E_iE_j$) and electric quadruple (EQ) term ($\chi^{EQ}_{ijkl}E_j\nabla_{k}E_l$). The rank 3 tensor $\chi^{ED}_{ijk}$ and rank 4 tensor $\chi^{ED}_{ijk}$ follow the symmetry of the MnPS$_3$ crystal. Here, we only consider the case that the mirror symmetry of MnPS$_3$ is present.
\subsection{$T>T_N$}
Above N\'eel temperature, the point group of MnPS$_3$ crystal is 2/m. The $\chi^{ED}_{ijk}E_iE_j$ is zero because of the presence of the inversion symmetry. Under normal incidence, we can simplify the EQ term as $\chi^{EQ}_{ijzl}k_zE_jE_l (j,l=x,y)$. According to the mirror symmetry perpendicular to the b-axis we have $\chi^{EQ}_{xxzy}=\chi^{EQ}_{xyzx}=\chi^{EQ}_{yxzx}=\chi^{EQ}_{yyzy}=0$. In conclusion, the SHG intensity for parallel and crossed configuration should be
\begin{equation}
    I_{parallel}(2\omega,T>T_N)\propto\left((\chi^{EQ}_{xyzy}+2\chi^{EQ}_{yxzy})\sin^2{\phi}\cos{\phi}+\chi^{EQ}_{xxzx}\cos^3{\phi}\right)^2I(\omega)^2
\end{equation}
and
\begin{equation}
    I_{crossed}(2\omega,T>T_N)\propto\left((-\chi^{EQ}_{xxzx}+2\chi^{EQ}_{yxzy})\sin{\phi}\cos^2{\phi}-\chi^{EQ}_{xyzy}\sin^3{\phi}\right)^2I(\omega)^2,
\end{equation}
where $I(\omega)$ is the incident power and $\phi$ is the angle between a-axis and polarization of incident light. When $\phi=\pi/2$, $I_{parallel}(2\omega)=0$, and when $\phi=0$, $I_{crossed}(2\omega)=0$, indicating no SHG output is from the b-axis.

\subsection{$T<T_N$}
When the long-range antiferromagnetic order forms, the MnPS$_3$ breaks inversion symmetry, so the $\chi^{ED}$ is non-zero. According to its magnetic point group 2'/m, we have $\chi^{ED}_{xxy}=\chi^{ED}_{xyx}=\chi^{ED}_{yxx}=\chi^{ED}_{yyy}=0$. Then the second-harmonic electric field for the parallel and crossed configurations from ED term are,
\begin{equation}
    E_{parallel}^{ED}(2\omega,T<T_N)\propto\left((\chi^{ED}_{xyy}+2\chi^{ED}_{yxy})\sin^2{\phi}\cos{\phi}+\chi^{ED}_{xxx}\cos^3{\phi}\right)I(\omega)
\end{equation}
and
\begin{equation}
    E_{crossed}^{ED}(2\omega,T<T_N)\propto\left((-\chi^{ED}_{xxx}+2\chi^{ED}_{yxy})\sin{\phi}\cos^2{\phi}-\chi^{ED}_{xyy}\sin^3{\phi}\right)I(\omega).
\end{equation}

The symmetry of the tensor $\chi_{ijkl}^{EQ}$ is identical to the case of $T<T_n$, so 
\begin{equation}
    E_{parallel}^{EQ}(2\omega,T<T_N)\propto\left((\chi^{EQ}_{xyzy}+2\chi^{EQ}_{yxzy})\sin^2{\phi}\cos{\phi}+\chi^{EQ}_{xxzx}\cos^3{\phi}\right)I(\omega)
\end{equation}
and
\begin{equation}
    E_{crossed}^{EQ}(2\omega,T<T_N)\propto\left((-\chi^{EQ}_{xxzx}+2\chi^{EQ}_{yxzy})\sin{\phi}\cos^2{\phi}-\chi^{EQ}_{xyzy}\sin^3{\phi}\right)I(\omega).
\end{equation}

Finally, the SHG intensity below $T_N$ is
\begin{equation}
    I_{parallel}^{total}(2\omega,T<T_N)\propto\left(\pm E_{parallel}^{ED}(2\omega,T<T_N)+E_{parallel}^{EQ}(2\omega,T<T_N)\right)^2
\end{equation}
and
\begin{equation}
    I_{crossed}^{total}(2\omega,T<T_N)\propto\left(\pm E_{crossed}^{ED}(2\omega,T<T_N)+E_{crossed}^{EQ}(2\omega,T<T_N)\right)^2,
\end{equation}
where $\pm$ corresponds to two different domains.

\begin{figure*}
\centering
\includegraphics[width=\textwidth]{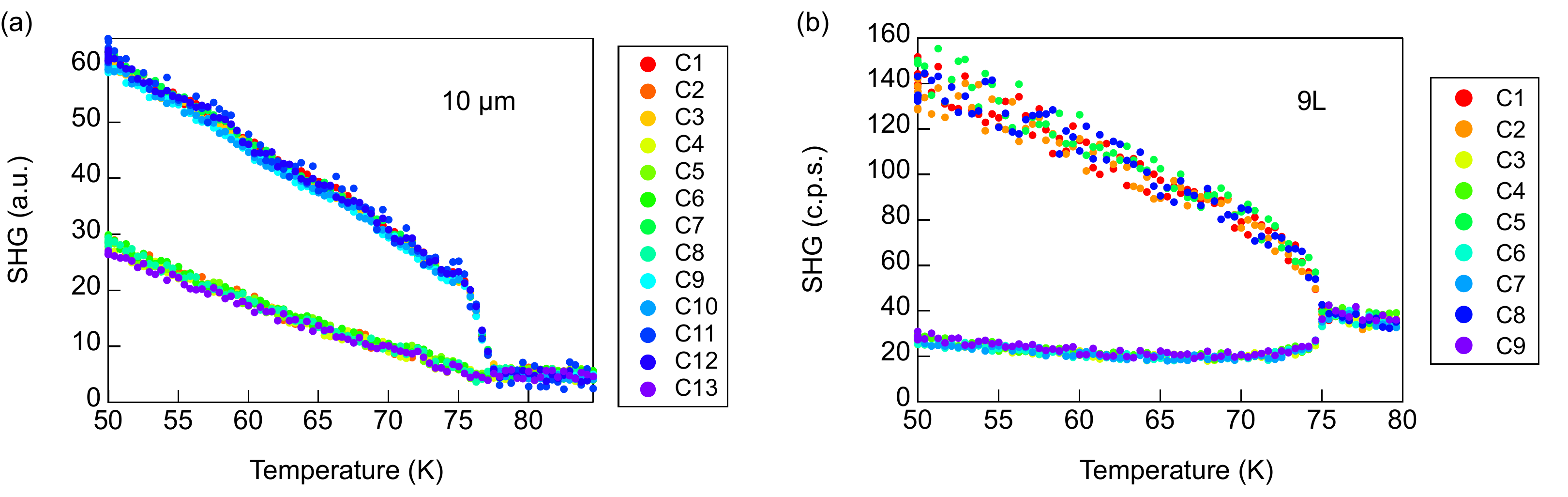}
\caption{\textbf{Consecutive temperature dependence of SHG measurement on MnPS$_3$.} (a) Cooling curves under 13 consecutive thermal cycles on the 10 $\mu$m bulk crystal shown in Fig.1 in the main text. (b) Cooling curves under 9 consecutive thermal cycles on the 9L sample shown in Fig.2 in the main text. In both cases, only two kinds of SHG signals are observed.}
\label{figS4}
\end{figure*}

\section{SHG measurements on NiPS$_3$}
\begin{figure}
\centering
\includegraphics[width=0.95\textwidth]{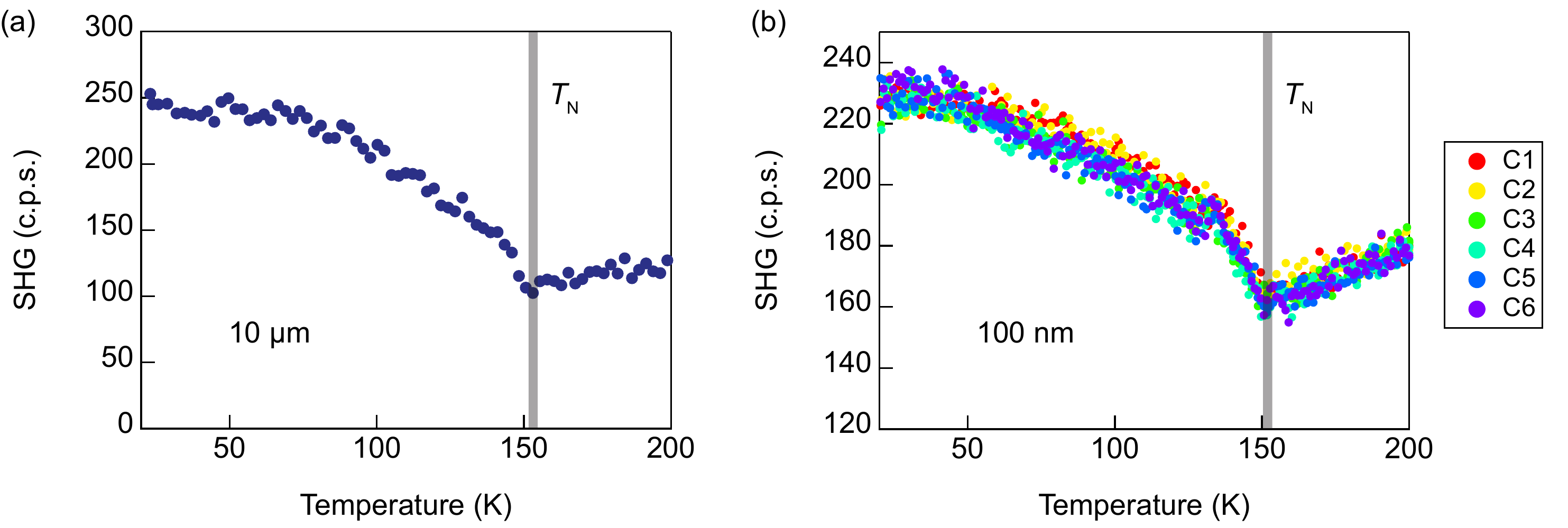}
\caption{\textbf{Temperature dependence of SHG intensity of NiPS$_3$.} (a) Temperature dependence of SHG intensity of a NiPS$_3$ bulk crystal of the thickness around 10 $\mu$m. A transition-like feature happens around 153 K, which is the antiferromagnetic transition temperature of NiPS$_3$. (b) Temperature dependence of SHG intensity of a thick NiPS$_3$ flake exfoliated on SiO$_2$/Si substrate. Six consecutive thermal cycles are performed to confirm the existence of only time-reversal invarient contribution.}
\label{nips3}
\end{figure}

Here we measure a centrosymmetric van der Waals antiferromagnetic material NiPS$_3$. The NiPS$_3$ crystal has the same space and point group in the paramagnetic phase as MnPS$_3$, but it is in a so-called ``zigzag'' AFM state under the transition temperature around 153 K. The zigzag AFM does not break inversion, so the spin-induced c-type SHG is not allowed in this compound. However, the i-type electric quadrupole SHG is allowed and influenced by the magneto-elastic coupling with the formation of the 
antiferromagnetic order. The temperature dependence of a NiPS$_3$ crystal with a thickness around 10 $\mu$m is shown in Figure \ref{nips3}(a). Power of 5 mW and a beam spot diameter of 10 $\mu$m are used. Note that a clear transition happens at the N\'eel temperature. We also measure a 100-nm thick NiPS$_3$ flake exfoliated on the SiO$_2$/Si wafer, and see a similar transition at around 153 K (Fig. \ref{nips3}(b)). Consecutive thermal cycles are performed to confirm that only one possible SHG signal exists below the N\'eel temperature, which is consistent with the absence of the spin-induced c-type SHG contribution. Note that SHG was reported without a change across the the N\'eel temperature in NiPS$_3$  \cite{chuprl20}. The observation of the phase transition due to magneto-elastic coupling also shows the higher sensitivity of our setup.

\section{Crossover of the exponent component $\beta$ in M\lowercase{n}PS$_3$ bulk}

\begin{figure}
\centering
\includegraphics[width=0.5\textwidth]{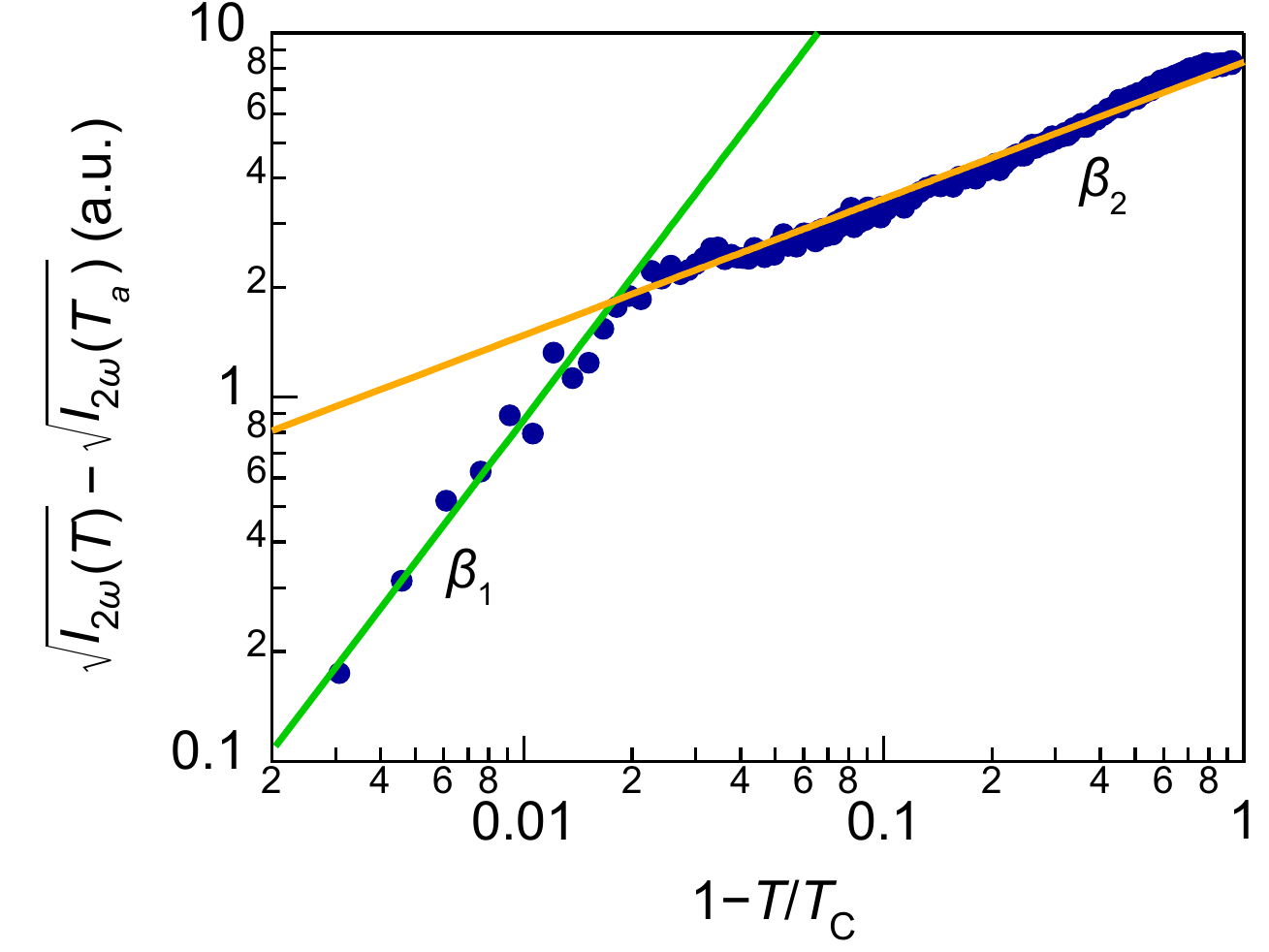}
\caption{\textbf{Temperature dependence of the  second-harmonic electric field below $T_N$ in the log-log plot.} The data (dark blue circles) are from the domain 1 of the 10 $\mu$m-thick MnPS$_3$ crystal in  Fig. 1 in the main text. A crossover of slope $\beta$ is observed at around 76 K. Linear fitting is performed above and below 76 K, as shown by the green and the yellow lines.}
\label{beta}
\end{figure}

Near the N\'eel temperature, the electric-dipole second-order susceptibility $\chi^{ED}$ is proportional to $(1-T/T_N)^{\beta}$. In Fig.1 in the main text, we report a change of slope around 76 K in the temperature-dependent SHG  in a bulk MnPS$_3$ crystal. To study the crossover-like feature, we plot the square root of SHG signal of domain 1 (the nonzero value at $T_N$ is subtracted) as a function of $1-T/T_N$ in the log-log plot (Fig. \ref{beta}). The slope of the curve is then representing the critical exponent $\beta$. A clear crossover of the slope is observed in the plot with $\beta_1=1.3(0)$ and $\beta_2=0.37(5)$.  A neutron scattering measurement in MnPS$_3$ also shows a crossover feature on the exponent coefficients at a similar temperature but with different values of $\beta$ \cite{wildesprb2006}, where the critical exponents are more accurate as it measures the staggered magnetization.  Note that $\chi^{ED}_{ijk}(T) \propto$ $M$ $\propto(T_N-T)^{\beta}$ \cite{muthukumarprl95}. If the EQ SHG does not depend on temperature below $T_N$, then we will see similar critical exponents as neutrom scattering.   Therefore, we conclude that electric-quadruple contribution is also dependent on temperature and can show a phase-transition-like behavior (See NiPS$_3$ data in Note S2). In this regard, the $\beta_1$ and $\beta_2$ in Fig. \ref{beta} are not purely related with $\chi^{ED}$ as it also include the EQ SHG that is also dependent on the temperature below $T_N$.

\section{Layer characterization of few-layer M\lowercase{n}PS$_3$}
\begin{figure}
\centering
\includegraphics[width=\textwidth]{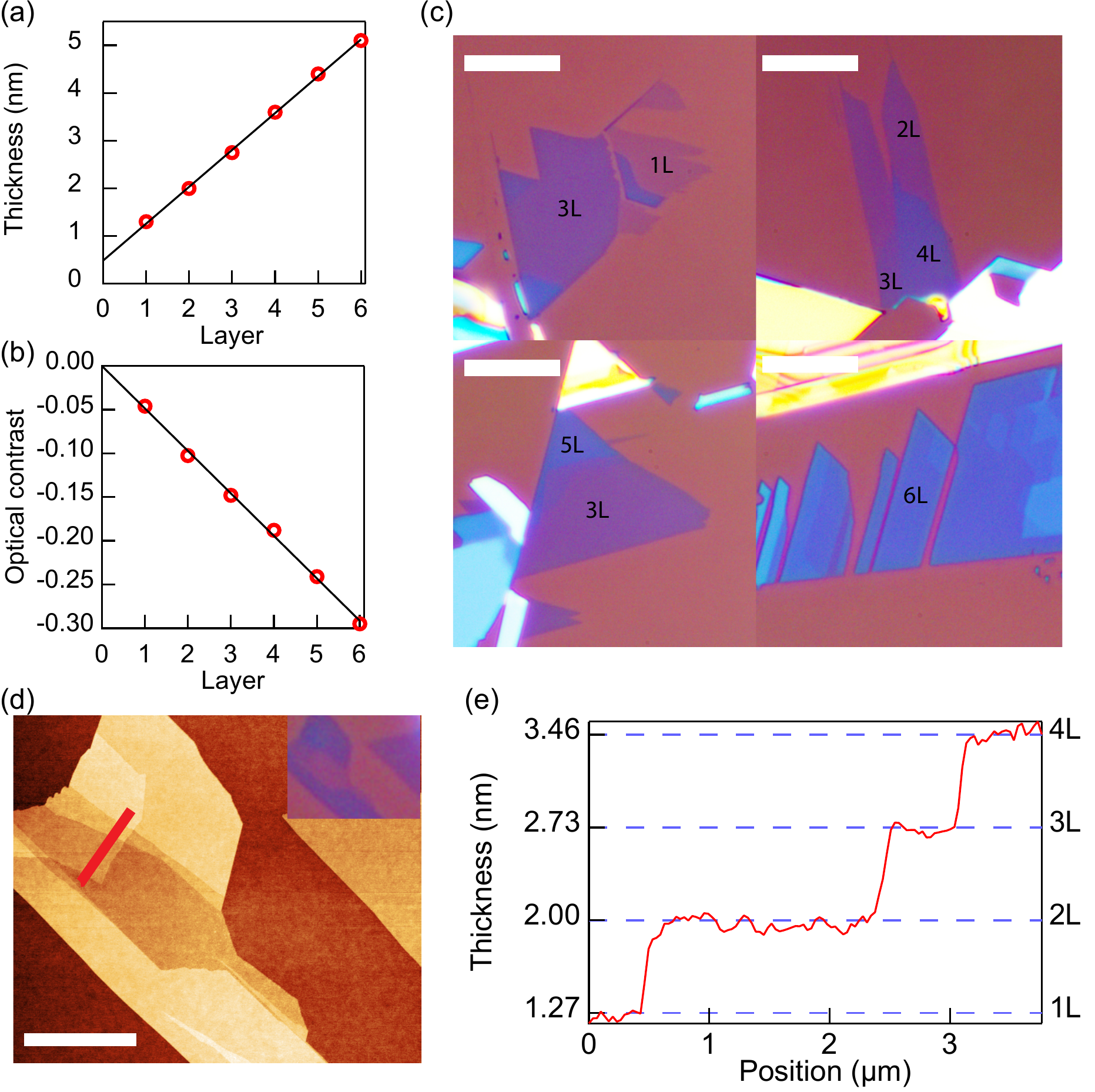}
\caption{\textbf{Determination of layer numbers.} (a) Layer-dependent thickness measured by atomic force microscopy. (b) Layer-dependent optical contrast measured determined by the red channel of the optical image. (c) Optical images of typical few-layer MnPS$_3$. Scale bar: 20 $\mu$m. The white balance is adjusted from the default setting to maximize the contrast of different layers. (d) Atomic force microscopy image of a multi-layer flake. (Inset: the optical image of the same flake). Scale bar: 5 $\mu$m. (e) A line cut of the red line shown in (d) with a single layer step of around 0.73 nm.}
\label{figS1}
\end{figure}

Typical optical images of different layers are shown in Fig. \ref{figS1}(a). We use two methods to distinguish the number of layers, including atomic force microscopy and optical contrast. For the atomic force microscopy measurement, the height information of the contact mode is used. A single-layer step of around 0.73 nm can be seen from Fig. \ref{figS1}(e). The height of the monolayer is measured to be around 1.3 nm, which is larger than a single-layer step. It is possible because of the existence of an air layer on the sample surface. Layer-dependent sample thickness is shown in Fig. \ref{figS1}(a) by open red circles, and a linear fit is marked by the black line. We also use optical contrast to determine the layer numbers. We extract the red channel of optical image and calculate the optical contrast for different samples by $\eta=\frac{I_{sample}-I_{substrate}}{I_{sample}+I_{substrate}}$. The layer-dependent optical contrast is plotted in Fig. \ref{figS1}(b) by open red circles with a linear fit marked by a black line.

\section{Extra data for the bilayer M\lowercase{n}PS$_3$ sample.}
\begin{figure*}
\centering
\includegraphics[width=\textwidth]{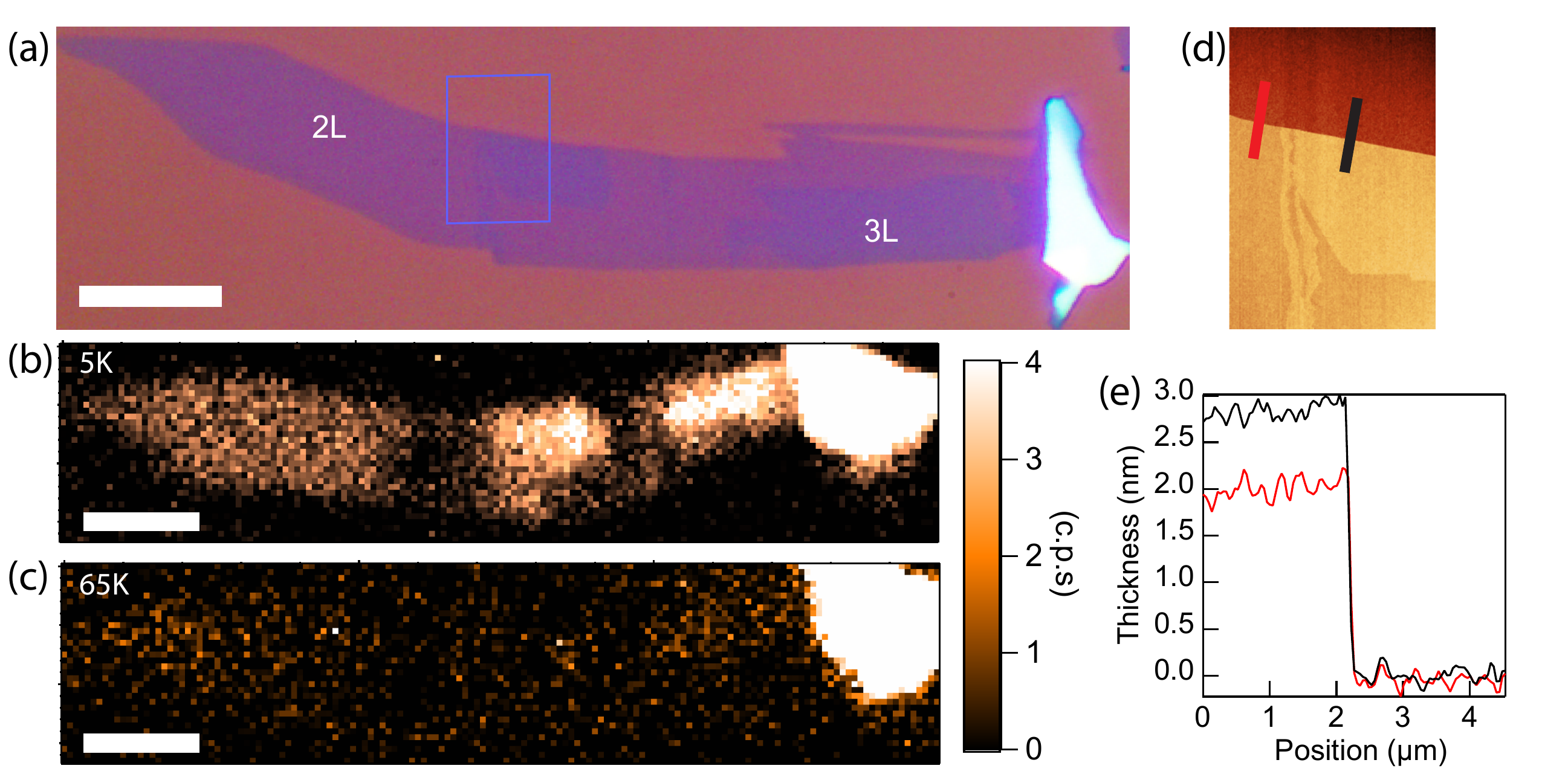}
\caption{\textbf{Extra data for the bilayer MnPS$_3$ sample in the main text.} (a) Optical image of the sample. Scale bar: 20 $\mu$m. (b) SHG intensity mapping at 5 K. Scale bar: 20 $\mu$m. (c) SHG intensity mapping measured at 65 K. Scale bar: 20 $\mu$m. (d) Atomic force microscopy height image of the boxed area in (a). The red and black cut lines are shown in (e) with a thickness step of 2.0 nm and 2.7 nm from the substrate, corresponding to a bilayer and a trilayer respectively.}
\label{figS2}
\end{figure*}

Here we show some extra data of the bilayer sample in the main text. The optical image, atomic force microscopy mapping, and SHG intensity mapping at 5 K and 65 K are shown in Fig. \ref{figS2}. Clearly, the SHG signal exists at 5 K while almost disappears at 65 K for the left-side bilayer area. The data shown in the main text was measured at the same spot in the middle of the left-side bilayer area.

\begin{figure*}
\centering
\includegraphics[width=\textwidth]{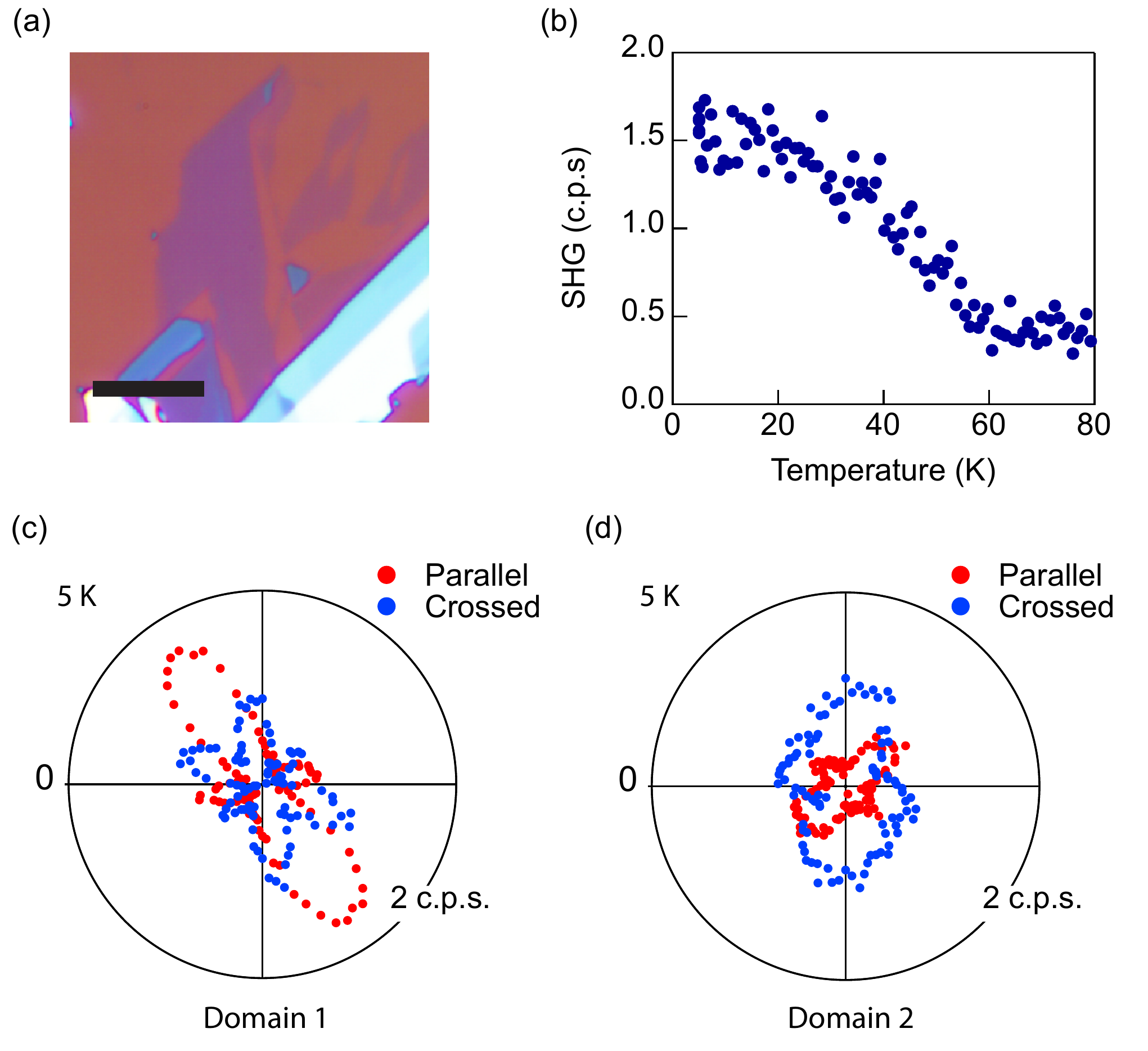}
\caption{\textbf{SHG data for a bilayer sample exfoliated in the glove box.} (a) Optical image of the sample. Scale bar: 20 $\mu$m. (b) SHG intensity as a function of temperature (c-d) Polarization dependent SHG patterns in domain 1 and domain 2.}
\label{bilayerglovebox}
\end{figure*}

To exclude the possibility of degradation of the bilayer sample during the sample preparation process, we also measured a bilayer sample S2 exfoliated in a glove box with the O$_2$ and water level less than 1 ppm. The data are shown in Fig. \ref{bilayerglovebox}. The transition temperature  extracted from Fig. \ref{bilayerglovebox} (b) is 61 K, and the SHG intensity is around 1.5 c.p.s.. These values are consistent with those from the bilayer sample exfoliated in air shown in the main text.

\section{Characterization of monolayer M\lowercase{n}PS$_3$ samples.}
\begin{figure*}
\centering
\includegraphics[width=\textwidth]{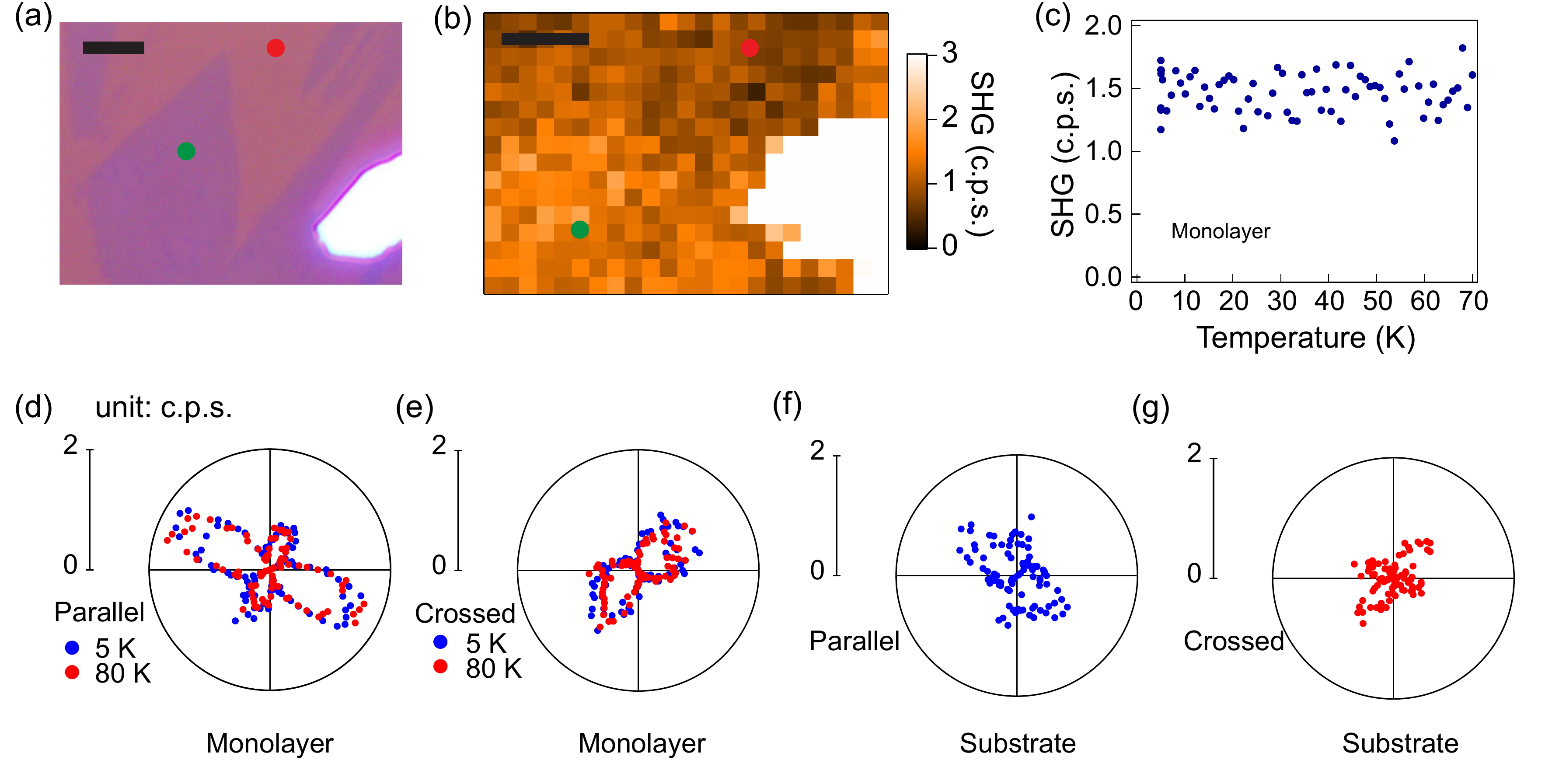}
\caption{\textbf{SHG data of the monolayer MnPS$_3$ exfoliated in a glove box.} (a) Optical image of the monolayer sample. Scale bar: 10 $\mu$m. (b) A SHG mapping of the monolayer sample. The region around the green dot is the monolayer and the region around the red dot is the SiO$_2$/Si substrate. Scale bar: 10 $\mu$m. (c) Temperature-dependent SHG signal. (d-e) Parallel (d) and crossed (e) patterns of the monolayer sample at 5 K and 80 K. (f-g) Parallel (f) and crossed (g) pattern of the substrate.}
\label{monolayer}
\end{figure*}

We measured the data from three different monolayer MnPS$_3$ samples, and did not observe a temperature-dependent SHG within the sensitivity of our setup. The data of the monolayer shown in the main text were measured under the laser power of 2.0 mW. The monolayer sample was exfoliated in air, and the optical images are shown in Fig.\ref{figS1}(c). A second sample prepared and measured under the same condition shows the same temperature-independent signal. We also tried one monolayer sample exfoliated in the glove box to avoid any possible degradation during the sample preparation process, though the MnPS$_3$ monolayer is reported to be stable in air \cite{leeaplm2016}. The details of this sample are shown in Fig. \ref{monolayer}. In this sample, we used a relatively high laser power of 5.0 mW, which generates enough photons to measure polarization-dependent SHG patterns. We observed a small SHG signal on top of the signal from the substrate, but we did not observe any temperature-dependent signals between 5 K and 80 K.

The long-range order in bulk MnPS$_3$ is believed to result from  the anisotropy and the weak interplanar coupling \cite{ronnowPhysicaB2000,wildesprb2006}. Except for the ordering near the 3D N\'eel temperature, the spin dynamics of a bulk sample are best described by the 2D Heisenberg Hamiltonian with  weak XY anisotropy and interlayer coupling \cite{wildesprb2006}. If atomically thin samples have the same Hamiltonian as the bulk, the absence of the long-rage order in the monolayer is either consistent with the 2D Heisenberg model without interlayer coupling or in the same universality class as the 2D XY model, which we do not have direct evidences for.  Nevertheless, as we wrote in the main text, both the lattice and the spin structure of samples thinner than 10L are found to be different from the bulk, so we actually do not know the spin Hamiltonian and anisotropy in atomically thin samples. Note that the suppression of $T_N$ is also most prominent below 10L.  It is possible that the anisotropy is strongly reduced in atomically thin limit, especially in the monolayer \cite{gongnat17,dengnat18}. Also, we can not rule out the possibility of disorder effect  even though we do not see buckling from atomic force microscopy measurements. We hope our work will motivate more works on atomically thin MnPS$_3$ especially the monolayer in the future.

\section{Strain-tuning measurement of  few-layer M\lowercase{n}PS$_3$ samples}
From the SHG measurement, we have confirmed that the mirror symmetry was broken below and above the N\'eel temperature in few-layer samples with thickless less than 11L. It is reported previously in a 7L MnPS$_3$ sample below N\'eel temperature by a similar SHG measurement, which they attribute to the substrate-induced strain\cite{chuprl20}. According to our measurement, we argue that the substrate-induced strain is not the key reason for the mirror-symmetry breaking. First, strain modification on SHG patterns is usually quite small. The polarization-dependent SHG patterns of the few-layer samples shown in the main text are quite different from the six-fold patterns in the thicker samples. While the previous study on strain-dependent SHG of other 2D materials never shows such a large modification even by adding a large strain deliberately\cite{Mennelnc18,mennelaplp19}.

\begin{figure*}
\centering
\includegraphics[width=\textwidth]{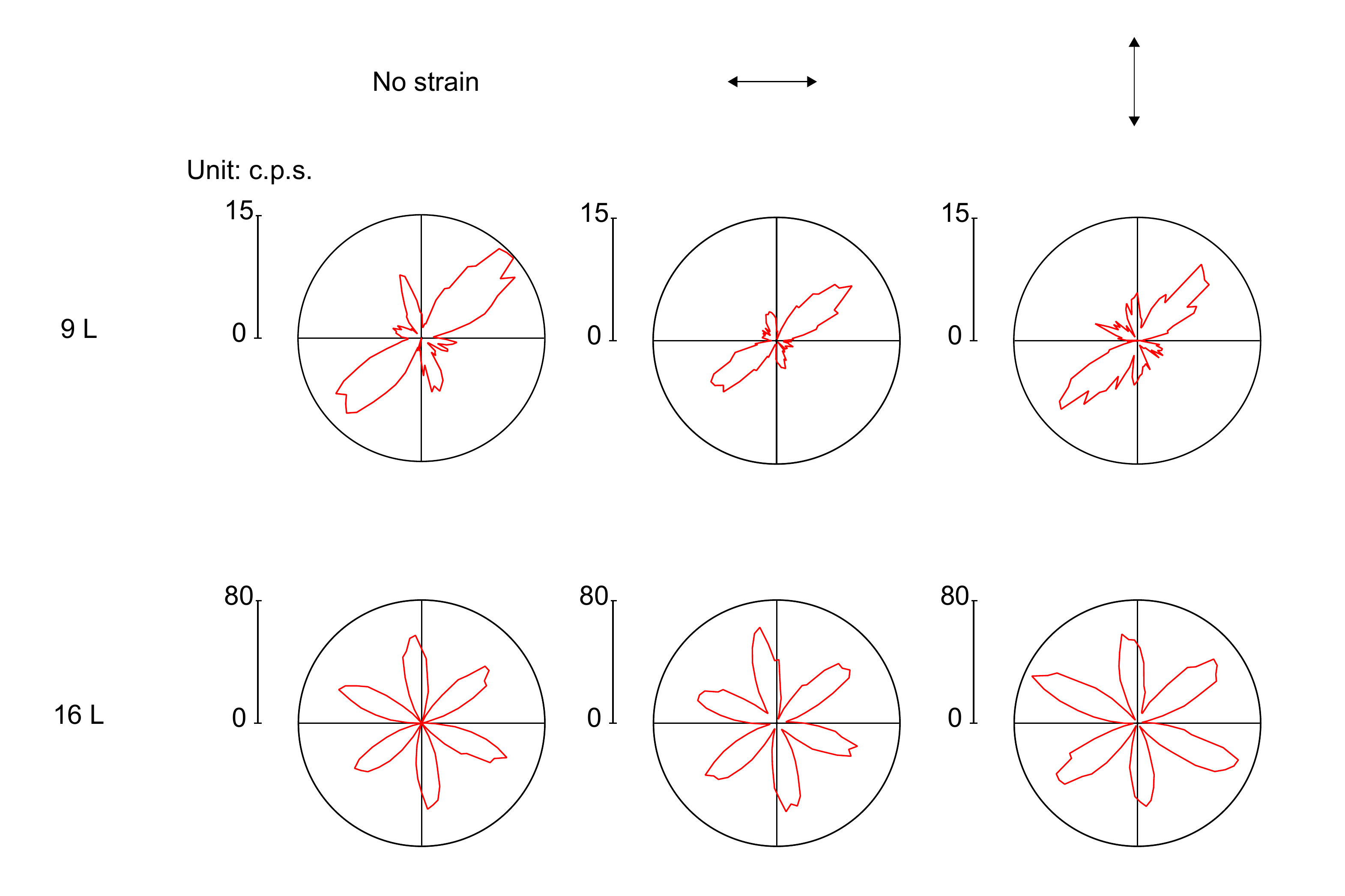}
\caption{\textbf{Crossed SHG patterns of samples on a substrate under uniaxial strain.}}
\label{strain}
\end{figure*}

To further justify our statement, we did a strain-tuning experiment on both thick and thin MnPS$_3$ samples. MnPS$_3$ were exfoliated on a polydimethylsiloxane (PDMS) substrate. The flexibility of the PDMS film enabled us to add strains on the samples in any direction\cite{liunatcomm14, zhangafm16,ninatnano21}. We used a micro-mechanical stage to apply around 5\% uniaxial strain on the PDMS, which is then attached to a gold-coated wafer. The transfer ratio of the strain is reported to be around 13-15\% \cite{liunatcomm14, zhangafm16,ninatnano21}.  The adherence between the gold coating and PDMS was strong enough to sustain the strain at 5 K. Fig.\ref{strain} shows the results of a 9L sample and a 16L sample measured at 5 K and under different  strains applied on the substrates.  We did not observe a clear difference between different strains in both samples. Specifically, no obvious mirror symmetry breaking was induced in the 16 L sample. 

\bibliography{2Dmag}